\documentclass{article}

\usepackage{microtype}
\usepackage{graphicx}
\usepackage{subfigure}
\usepackage{booktabs} 

\usepackage{hyperref}

\usepackage{optidef}
\usepackage{scalerel}
\graphicspath{{images}}
\usepackage{siunitx}

\usepackage[accepted]{icml2025}

\usepackage{amsmath}
\usepackage{amssymb}
\usepackage{mathtools}
\usepackage{amsthm}

\usepackage{booktabs} 
\usepackage{multirow} 
\usepackage{array}    

\usepackage{amsfonts}

\usepackage[capitalize,noabbrev]{cleveref}

\theoremstyle{plain}

\theoremstyle{definition}

\theoremstyle{remark}

\usepackage[textsize=tiny]{todonotes}

\icmltitlerunning{Real-Time Object Tracking with On-Device Deep Learning for Adaptive Beamforming in Dynamic Acoustic Environments}

\begin{document}

\twocolumn[
\icmltitle{Real-Time Object Tracking with On-Device Deep Learning for Adaptive Beamforming in Dynamic Acoustic Environments}



\icmlsetsymbol{equal}{*}

\begin{icmlauthorlist}
\icmlauthor{Jorge Ortigoso-Narro}{aaa}
\icmlauthor{Jose A. Belloch}{aaa}
\icmlauthor{Adrian Amor-Martin}{bbb}
\icmlauthor{Sandra Roger}{ccc}
\icmlauthor{Maximo Cobos}{ccc}
\end{icmlauthorlist}

\icmlaffiliation{aaa}{Department of Electronic Technology, Universidad Carlos III de Madrid, Spain}
\icmlaffiliation{bbb}{Department of Signal Theory and Communications, Universidad Carlos III de Madrid, Spain}
\icmlaffiliation{ccc}{Department of Computer Science, Universidad de Valencia, Spain}

\icmlcorrespondingauthor{Jorge Ortigoso-Narro}{jortigos@pa.uc3m.es}

\icmlkeywords{Beamforming, Autograd, Concentric Circular Arrays, Frequency Invariant.}

\vskip 0.3in
]



\printAffiliationsAndNotice{Preprint, sumbitted to Journal of Supercomputing.}  

\begin{abstract}
Advances in object tracking and acoustic beamforming are driving new capabilities in surveillance, human-computer interaction, and robotics. This work presents an embedded system that integrates deep learning–based tracking with beamforming to achieve precise sound source localization and directional audio capture in dynamic environments. The approach combines single-camera depth estimation and stereo vision to enable accurate 3D localization of moving objects. A planar concentric circular microphone array constructed with MEMS microphones provides a compact, energy-efficient platform supporting 2D beam steering across azimuth and elevation. Real-time tracking outputs continuously adapt the array’s focus, synchronizing the acoustic response with the target’s position. By uniting learned spatial awareness with dynamic steering, the system maintains robust performance in the presence of multiple or moving sources. Experimental evaluation demonstrates significant gains in signal-to-interference ratio, making the design well-suited for teleconferencing, smart home devices, and assistive technologies.
\end{abstract}

\section{Introduction}
\label{sec:introduction}

The increasing demand for intelligent systems that can reliably localize sound sources and capture directional audio has made object tracking and acoustic beamforming essential technologies. Applications range from consumer-grade solutions such as teleconferencing, smart home automation, and assistive devices to mission-critical systems including anti-drone defense, perimeter surveillance, and autonomous security platforms. These systems rely on the integration of spatial awareness and audio enhancement to operate effectively in complex and dynamic environments~\cite{surveillance, Price99}. 

Object tracking and computer vision systems based on deep learning have become widely studied and substantially advanced in recent years~\cite{objectdetectionsurvey}. These methods deliver fast and reliable performance, enabling precise identification and spatial localization of objects, along with robust classification capabilities. Compared to audio-only systems, this enables a more intelligent, context-aware approach, enhancing situational awareness and interaction with dynamic environments. However, a key drawback is that the inference part of deep learning models is often computationally demanding, requiring specialized hardware such as Graphics Processing Units (GPUs) or dedicated Artifical Intelligence (AI) accelerators. Achieving real-time performance in such systems necessitates careful optimization and efficient model deployment, particularly in resource-constrained and portable platforms.

To enable real-time performance of deep learning–based computer vision models, a range of optimization strategies is employed. Model quantization, for instance, reduces weights and activations from floating-point to lower-precision formats (e.g., INT8), which decreases memory usage and computation time with minimal accuracy loss~\cite{jacob2017}. Pruning and architecture simplification further reduce inference latency by eliminating redundant parameters and layers~\cite{han2015, howard2017} improving the performance (specially in edge AI platforms).

In a complementary domain, Micro-Electromechanical Systems (MEMS) microphone arrays provide an efficient foundation for embedded audio systems due to their compact form factor, low power requirements, and ease of integration~\cite{ortigoso-narro_64-microphone_2024}. Their small size enables dense array configurations, which in turn support accurate beamforming and spatial localization in constrained environments. These arrays are increasingly adopted in robotics, smart assistants, and surveillance, where precise directional audio capture is required. Combined with modern signal processing hardware, they offer a practical solution for real-time spatial audio processing on embedded platforms.

The integration of visual tracking with acoustic beamforming has helped advance sound source localization, allowing for highly precise directional audio capture in both traditional and modern systems. Multimodal devices such as the Microsoft Kinect and Intel RealSense D455 combine RGB cameras, depth sensors, and microphone arrays to spatially focus audio by targeting a speaker’s mouth through 3D depth mapping, effectively reducing background noise. Recent developments, such as the work by Nagasha et al.~\cite{nagasha}, which merges face tracking with beamforming, and mobile robots addressing the cocktail party problem~\cite{cocktail}, leverage deep learning to improve performance in multi-speaker environments. 

The main contribution of this work is the design and implementation of a compact, energy-efficient embedded system that integrates deep learning–based 3D object localization with dynamically steerable MEMS microphone beamforming, enabling real-time directional audio capture in dynamic acoustic environments. Thus, we integrate hardware and software changes to improve the overall performance.

\section{Proposed system}

\begin{figure}[ht]
    \centering
    \includegraphics[width=0.8\linewidth]{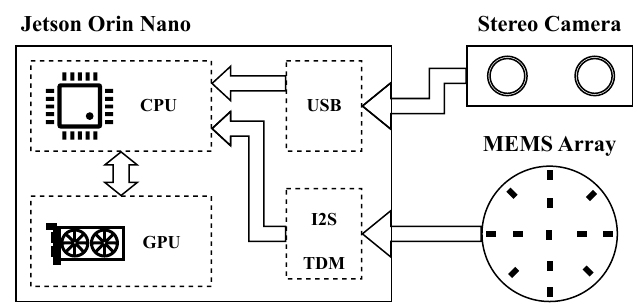}
    \caption{High-level block diagram design.}
    \label{fig:block_diagram}
\end{figure}

The proposed system consists of three primary components: (1) a visual perception module with depth estimation capabilities, utilizing either a stereo or monocular camera, (2) a planar concentric circular MEMS microphone array for audio acquisition, and (3) an NVIDIA Jetson Orin Nano (8GB version) embedded processor that handles tracking, classification, and beam steering. These components are integrated through a low-latency pipeline that facilitates closed-loop interaction between the visual and acoustic subsystems. Figure~\ref{fig:block_diagram} illustrates the high-level block diagram of the proposed system architecture.

\subsection{Vision Module}

To provide visual information to the system, a low-cost USB stereo camera (GXIVISION-LSM22100) was selected as the sensing hardware. The vision module therefore consists of two synchronized image sensors integrated in a single stereo unit. The joint resolution of the stereo pair is $2560\times720$ pixels, i.e., each camera provides frames at a resolution of $1280\times720$ pixels. The camera streams MJPEG-compressed video at $30$~FPS over USB, which offers a cost-effective solution while still providing sufficient spatial resolution and frame rate for real-time object tracking and depth estimation.

\subsubsection{Object classification}

Object classification has been a key focus in computer vision research. Early approaches leaned on carefully designed features, like Scale Invariant Feature Transform (SIFT) and Histograms of Oriented Gradients (HOG), paired with simple classifiers such as Support Vector Machines~\cite{Lowe2004, dalal, Csurka2002VisualCW}. These methods worked well in specific, controlled environments but often faltered when faced with challenges like changing light conditions, partial object obstruction, or busy backgrounds.

With the rise of deep learning, Convolutional Neural Networks (CNNs) have taken center stage, delivering impressive gains in both accuracy and reliability across a wide range of image datasets. Models like AlexNet, VGG, and ResNet stand out for their ability to automatically learn layered patterns from raw images, eliminating the need for handpicked features. These architectures paved the way for more advanced systems, and architectures adapting to diverse real-world scenarios with greater ease~\cite{resnet, vgg}.

In recent advancements, the YOLO (You Only Look Once) series has gained traction for its efficiency in object classification and detection. Renowned for its rapid processing and dependable performance, YOLO supports real-time applications while maintaining strong accuracy. For our research, we selected YOLOv11, which refines earlier versions with sharper architectural enhancements, such as the C3k2 block and C2PSA attention mechanism, alongside optimized training techniques. These improvements enable YOLOv11 to reduce the number of parameters and maintain fast inference times. This model delivers a well-balanced solution, ensuring swift computation and consistent outcomes for our classification tasks~\cite{yolo11}.

\subsubsection{Depth Estimation}

To correctly estimate the direction of arrival of each source from the visual data, its 3D position in space is required. Depth estimation plays a central role in this process by providing the missing third dimension (labeled in this work as z-axis) to the 2D image-plane coordinates extracted from object detection.

One common approach to estimating depth is through stereopsis, which leverages a calibrated stereo camera setup with a known baseline $B$ and focal length
$f$. By identifying corresponding points in the left and right image frames, one can compute the disparity $d$, defined as the horizontal shift in pixels between the two views. The depth $\hat{z}$ of a point in the scene is then estimated using the well-known triangulation formula,
\begin{equation}
    \hat{z} = \frac{f \cdot B}{d}.
\end{equation}
 This method is geometrically grounded and yields accurate depth estimates when the stereo pair has sufficient baseline and reliable disparity computation. However, it can struggle in low-texture regions, occlusions, or repetitive patterns, where correspondence matching becomes ambiguous. Depth accuracy also decreases with distance due to reduced disparity.

An alternative involves deep learning-based approaches, including monocular and stereo depth estimation using deep neural networks~\cite{depth_anything_v2, monodepth2, rtmonodepth, lipson2021raft, li2022practical, XStereoLab2021}. Monocular methods predict depth from a single image using learned priors about object size, scene geometry, and perspective. These models, typically based on CNNs or transformer architectures, produce dense depth maps even in visually challenging scenes where stereo may fail although it comes with a cost: they are more computationally intensive and are useful in scenarios where we need to process a single image and hardware acceleration is available. 

In our case, where additional per-image processing is required, monocular depth estimation may reduce overall system latency and offer practical benefits despite its inherent limitations.

\subsubsection{Source Tracking}

After detecting and localizing objects and potential sound sources in the image, each instance is assigned a temporary identifier to maintain short-term consistency across frames. Note that a dedicated multi-object tracking algorithm is not used in this work since it offers minimal practical benefits: the system is designed for scenarios where overlapping objects have little influence on overall performance since beamforming cannot effectively separate coincident acoustic sources. 
Omitting this stage also reduces computational overhead and latency, resulting in a streamlined and responsive pipeline suitable for real-time embedded operation.

YOLO-based networks output bounding boxes for each detected object~$i$, characterized by their center coordinates $\mathbf{c}_i = (x_i, y_i)$, height~$h_i$, and width~$w_i$. In this work, we estimate the direction of arrival (DoA) of each source based on its visual centroid, which can be transformed into world coordinates using the camera calibration parameters.

Given the intrinsic camera matrix~$\mathbf{K}$ and the extrinsic parameters (rotation matrix~$\mathbf{R}$ and translation vector~$\mathbf{t}$), the image coordinates~$\mathbf{c}_i$ can be back-projected into a 3D ray in the camera coordinate system~\cite{Hartley2004}. Employing the known depth~$\hat{z}_i$ (from stereo vision or monocular deep learning-based depth estimation), the 2D centroid~$\mathbf{c}_i$ is first converted to normalized image coordinates~$\tilde{\mathbf{c}}_i$, 
\begin{equation}
    \tilde{\mathbf{c}}_i = \mathbf{K}^{-1}
    \begin{bmatrix}
    x_i \\
    y_i \\
    1
    \end{bmatrix}.    
\end{equation}
Then, the 3D position of the object in the camera frame is
\begin{equation}
    \mathbf{p}_i^{\text{cam}} = \hat{z}_i \cdot \tilde{\mathbf{c}}_i  .  
\end{equation}
Finally, the position in the world coordinate system is obtained by
\begin{equation}
    \mathbf{p}_i^{\text{world}} = \mathbf{R} \cdot \mathbf{p}_i^{\text{cam}} + \mathbf{t}.
\end{equation}

\begin{figure}[!htpb]
    \centering    
    \includegraphics[width=0.8\linewidth]{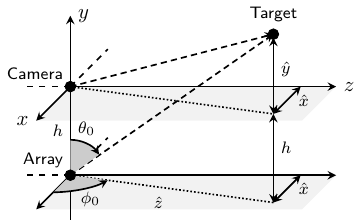} 
     \caption{System geometry for DoA estimation.}  
    \label{fig:geometry}
\end{figure}

This 3D position is used to estimate the direction of arrival by computing the azimuth and elevation angles with respect to the microphone array. To calculate the steering vector and beamforming coefficients for the array, the elevation angle $\theta_0$ and azimuth angle $\phi_0$ are derived from the spatial position of the detected target with 
\begin{equation}
    \phi_0 = \pi - \arctan \left(\frac{\hat{x}}{\hat{z}}\right) \quad
    \theta_0 = \arctan \left(\frac{\hat{y} + h}{\hat{z}}\right), \label{eq:theta_0}  
\end{equation}

where $h$ represents the vertical offset between the camera and the microphone array origins, assuming they are positioned on the same axis, as illustrated in Fig.~\ref{fig:geometry}, which shows the geometric model of the system and all relevant parameters.

\subsection{Beamforming Module}

The beamforming module serves as the core signal processing component responsible for spatially filtering the acoustic field captured by the microphone array. It utilizes directional information provided by the vision module to dynamically steer the beam.

\subsubsection{Frequency-Domain Delay-and-Sum Beamformer}

The delay-and-sum beamformer coherently combines signals from a microphone array to enhance sound arriving from a desired direction while suppressing off-axis sources. Given $M$ input channels $x_m[n]$ ($m = 1, \dots, M$), each microphone signal is time-shifted according to a steering delay $\tau_m$ that compensates for the propagation delay of a source from the target DoA. The beamformed output is thus obtained as
\begin{equation}
    y[n] = \frac{1}{M} \sum_{m=1}^{M} x_m[n + \tau_m f_s],
\end{equation}
where $f_s$ denotes the sampling frequency.

\paragraph{Steering delay computation.}
Let $\mathbf{r}_m = [x_m, y_m, z_m]^T$ denote the position vector of the $m$-th microphone in the array coordinate frame, and let $\mathbf{u}(\theta, \phi)$ be a unit vector pointing toward the target DoA, parameterized by azimuth $\phi$ and elevation $\theta$. Assuming a plane wave propagation model with sound speed $c$, the relative time delay for the $m$-th microphone is given by
\begin{equation}
    \tau_m = -\frac{\mathbf{r}_m^\top \mathbf{u}(\theta, \phi)}{c}.
    \label{eq:tau}
\end{equation}
The negative sign aligns the phases such that signals from the desired direction add constructively at the beamformer output.

\paragraph{Frequency-domain formulation.}
In the frequency domain, these time shifts correspond to complex phase rotations. Taking the discrete Fourier transform (DFT) of each channel signal yields
\begin{equation}
    X_m[k] = \sum_{n=0}^{N-1} x_m[n] \, e^{-j 2\pi kn / N},
\end{equation}
where $N$ is the frame length and $k$ is the frequency bin index.  
The beamformer output in the spectral domain is then computed as
\begin{equation}
    Y[k] = \frac{1}{M} \sum_{m=1}^{M} X_m[k] \, e^{j 2\pi f_k \tau_m},
    \label{eq:freq_beamform}
\end{equation}
with $f_k = k f_s / N$. The phase term $e^{j 2\pi f_k \tau_m}$ aligns the frequency components of all microphones toward the steering direction. The final beamformed signal is reconstructed by applying the inverse DFT:
\begin{equation}
    y[n] = \frac{1}{N} \sum_{k=0}^{N-1} Y[k] \, e^{j 2\pi kn / N}.
\end{equation}

Equation~\ref{eq:freq_beamform}, together with the geometric delay definition in Eq.~\ref{eq:tau}, fully specifies the frequency-domain delay-and-sum beamforming algorithm. 

\subsubsection{Hardware and Microphone Array Desing}

\begin{figure}[!htpb]
    \centering    
    \includegraphics[width=6.8cm]{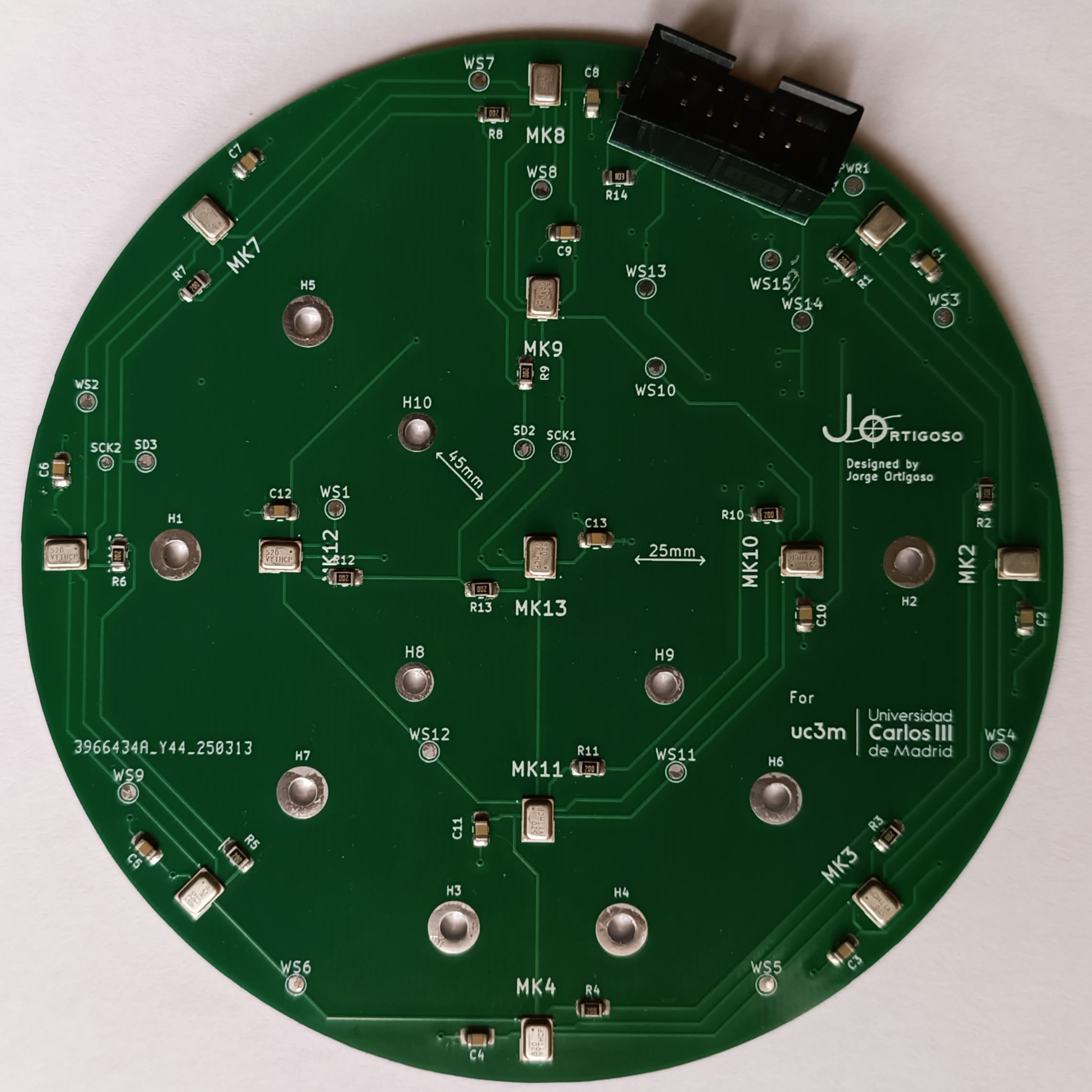} 
     \caption{Designed PCB containing the MEMS microphones. The sensors are placed along $R=3$ rings of radii $\rho = \{0, 2.5, 4.5\}\,$cm and equally spaced along each ring: for $r=2$, \ $\Delta\phi = \pi/2\,$ rad and for $r=3$, \ $\Delta\phi = \pi/4\,$ rad.}  
    \label{fig:pcb}
\end{figure}

A concentric circular microphone array was selected for its geometric symmetry, which supports flexible beam steering and consistent frequency response. This layout is well-suited for applications that require omnidirectional sound capture and spatial filtering, as it offers uniform angular resolution across the horizontal plane (in directions with the same number of microphones). The concentric design also allows the array to perform reliably over a broad frequency range, making it appropriate for sound source localization, speech enhancement, and acoustic scene analysis. The implemented array, illustrated in Fig.~\ref{fig:pcb}, consists of two circular rings with radii of $\SI{2.5}{\cm}$ and $\SI{4.5}{\cm}$, along with a central microphone.

The array was constructed using ICS-52000 MEMS microphones, which support daisy-chain operation. This feature consolidates the outputs of all microphones into a single digital stream, thereby simplifying data acquisition. A modified Inter-IC Sound (I\textsuperscript{2}S) interface was implemented, employing time-division multiplexing and a synchronization pulse to align data from each microphone correctly. This configuration provides reliable, low-latency, and synchronized audio transmission.

The printed circuit board (PCB) hosting the microphones was designed with a focus on signal integrity and timing accuracy. Termination resistors were placed adjacent to the microphone data pins (\SI{20}{\ohm}), while signal traces were routed with controlled impedance to match the system clock frequency. To minimize propagation delays, the data and clock lines were arranged in a branch-style topology, and signal buffers were incorporated to ensure adequate drive strength for both the microphone chain and the host interface.

\section{Optimizing for Real-Time Operation}

Achieving real-time performance in multimodal perception systems requires a unified approach to computational efficiency, data flow design, and hardware-aware optimization. The proposed system fully exploits the heterogeneous capabilities of the NVIDIA Jetson Orin Nano, where parallel execution across CPU and GPU resources enables synchronized processing of visual and acoustic streams with minimal latency. The architecture prioritizes throughput, determinism, and modularity to ensure stable operation under dynamic, real-world conditions.

\subsection{Parallel Multimodal Processing}

\begin{figure*}[!htpb]
    \centering    
    \includegraphics[width=\linewidth]{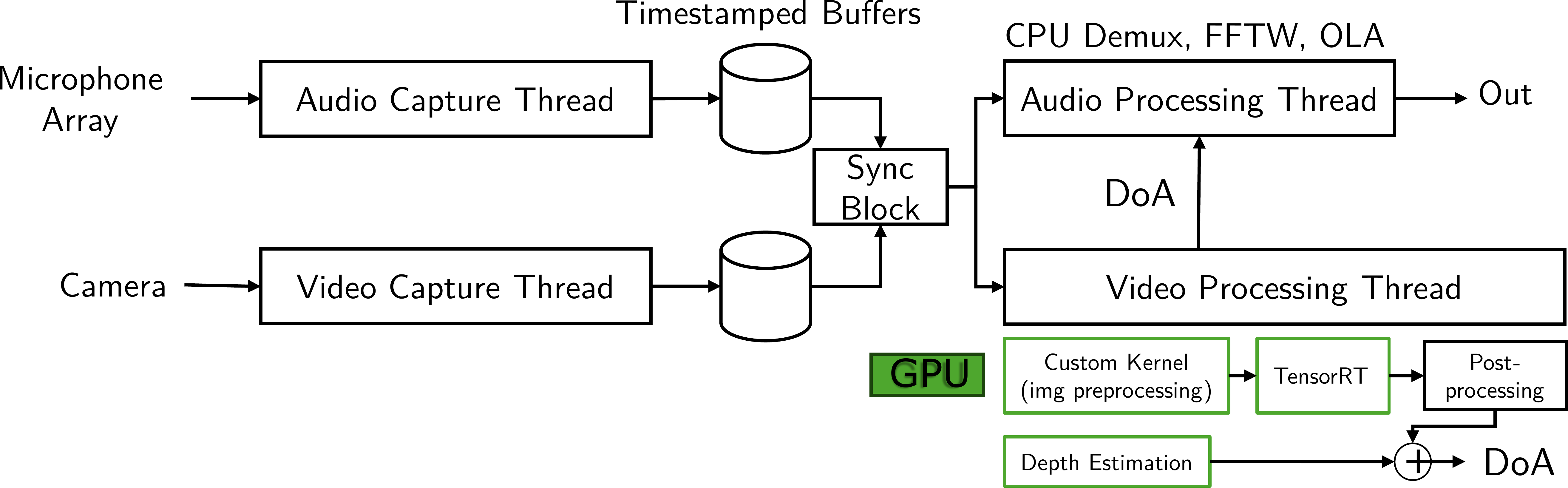} 
     \caption{Implemented thread diagram and acquisition/processing pipeline. For acronym definition see Sec.~\ref{subsec:overlapadd}.}  
    \label{fig:thread_diagram}
\end{figure*}

The real-time architecture (see Fig.~\ref{fig:thread_diagram}) adopts a modular pipeline that decomposes processing into concurrent subsystems for visual perception,  acoustic capture, beamforming, and control. Each module operates as an independent  thread, communicating through producer-consumer message queues implemented as  shared-memory deques protected by mutexes and condition variables. Communication  occurs exactly when a capture thread produces new data: the video-capture thread  pushes each GPU frame and timestamp into its queue and signals the video-processing  thread, while the audio-capture thread pushes each interleaved sample block with  timing metadata into the audio queue and wakes the beamformer. Consumer threads  block on their queues and resume processing immediately when notified. This  organization minimizes blocking dependencies and ensures that computationally  intensive tasks such as object detection never delay time-critical audio capture.

A timestamp-based synchronization scheme aligns multimodal data without rigid frame coupling.Each video frame and audio chunk is tagged using a common reference clock. During  fusion, the beamformer computes the midpoint timestamp of each audio chunk and retrieves  from the DOA history the latest visual estimate whose timestamp is not newer than that  midpoint. This closest-not-future criterion selects the most temporally relevant  sample available without waiting for future video frames, enabling continuous tracking  and beam steering with minimal buffering. This asynchronous, timestamp-driven approach follows common practice in real-time robotic perception pipelines, where decoupled processing threads reduce latency by preventing blocking due to different acquisition rates \cite{Quigley2009ROSAO, Qian2025}.

\subsection{GPU-Accelerated Perception and Frequency-Domain Beamforming}

Visual perception tasks are executed on the embedded GPU to leverage its high parallel throughput and minimize CPU load. Following the approach described previously, the YOLOv11-Nano architecture is employed for object detection, chosen for its balance between inference accuracy and computational efficiency on embedded hardware. The resulting 3D coordinates from object detection are transferred to the CPU as compact metadata for integration with the beamforming controller.

To further optimize the visual front-end, all camera preprocessing operations, including undistortion and input normalization, are executed directly on the GPU. The undistortion step leverages precomputed camera calibration parameters to correct lens distortion before inference, ensuring geometric consistency for accurate spatial mapping without requiring CPU intervention. Following undistortion, a custom CUDA kernel performs fused image preprocessing for the neural network input. This kernel applies an affine warp to scale and center the frame, performs bilinear interpolation for subpixel accuracy, normalizes pixel intensities, and reorganizes the image layout from interleaved to planar channel format.

Each CUDA thread processes one output pixel, computing its corresponding source coordinates via an affine transform and performing bilinear interpolation over the neighboring source pixels. Out-of-bound pixels are assigned a constant background value to preserve memory safety. The kernel uses a stride-based indexing approach to efficiently handle varying image widths and employs coalesced memory access for high throughput.

At the same time, the acoustic subsystem performs frequency-domain delay-and-sum beamforming to enable flexible and high-resolution spatial filtering. Each audio frame undergoes a short-time Fourier transform (STFT), delay compensation, and inverse transformation back to the time domain. 

\subsection{Overlap–Add Beamforming Implementation}
\label{subsec:overlapadd}

To ensure continuity between processed audio segments and to avoid artifacts introduced by frame-wise processing, the beamforming module employs an \textit{overlap–add} (OLA) strategy for frame reconstruction. Each input chunk of length $N=256$ is processed with a hop size of $H = N/2$, providing 50\% overlap between successive frames. This method maintains temporal smoothness and minimizes spectral discontinuities at chunk boundaries, crucial for preserving signal integrity in real-time streaming.

For each frame, multi-channel input signals $x_m[n]$ are transformed to the frequency domain using the discrete Fourier transform (DFT). The phase alignment and summation described in Eq.~\ref{eq:freq_beamform} are then performed for each frequency bin, followed by an inverse DFT to recover the time-domain frame $y[n]$.

The output frames are reconstructed through OLA accumulation,
\begin{equation}
    y_{\text{OLA}}[n] = \sum_{r} \left( y_r[n - rH] \cdot w[n - rH] \right),
\end{equation}
where $y_r[n]$ denotes the $r$-th frame and $w[n]$ is the synthesis window (a Hann window in our case). This ensures constant overlap gain and minimizes edge effects.

In practice, the FFT and IFFT operations are implemented using the FFTW library~\cite{fftw} on the CPU, which was empirically found to outperform GPU-based transforms for small frame sizes due to reduced data transfer overhead and superior cache utilization. Profiling results indicate that the dominant latency in the processing pipeline originates from visual inference rather than beamforming, making CPU-based spectral processing optimal for maintaining real-time responsiveness.

\section{Experiments and Results}

A series of experiments is conducted to evaluate the performance and real-time behavior of the complete system. The first stage focuses on comparing several depth-estimation methods in terms of computational efficiency and suitability for embedded operation. The second stage examines the full multimodal pipeline, combining vision-based localization and beamforming under both controlled and dynamic acoustic conditions. To minimize processing latency, the audio sampling rate was fixed at \SI{8}{\kHz}, balancing temporal resolution with computational load.  

The system was tested in two complementary scenarios: an anechoic environment to establish baseline performance and a generic room to assess robustness under realistic conditions.

\subsection{Latency Evaluation}

System latency was evaluated with all modules (visual inference, depth estimation, and beamforming) running concurrently to replicate real deployment conditions. This ensured that the measurements reflect the behavior of the integrated system, including synchronization overhead and GPU scheduling contention. Profiling combined direct timing instrumentation with advanced GPU tracing using NVIDIA Nsight Systems to characterize workload distribution and identify potential bottlenecks.

Table~\ref{tab:depth_latency} presents the average execution times of the evaluated depth estimation methods during full-system operation. Classical stereo approaches, such as filtered SGBM and the learning-based CREStereo, required longer computation times but offered the advantage of producing true disparity measurements. In contrast, lightweight monocular deep learning methods like Depth-AnythingV2 (metric) and RT-Mono-Depth achieved inference times below 0.2~s per frame, making real-time operation on embedded hardware more feasible.

\begin{table}[!ht]
    \centering
    \caption{Average execution time of evaluated depth estimation methods (mean~$\pm$~standard deviation).}
    \label{tab:depth_latency}
    \resizebox{\columnwidth}{!}{%
    \begin{tabular}{lc}
        \hline
        \textbf{Method} & \textbf{Latency (s)} \\
        \hline
        CREStereo~\cite{li2022practical} & $1.71 \pm 0.45$ \\
        RAFT-Stereo~\cite{lipson2021raft} & $5.67 \pm 1.07$ \\
        StereoNet~\cite{XStereoLab2021} & $1.26 \pm 0.39$ \\
        Depth-Pro~\cite{Bochkovskii2024} & $32.32 \pm 11.44$ \\
        Depth-AnythingV2~\cite{depth_anything_v2} & $1.33 \pm 0.20$ \\
        Depth-AnythingV2 (metric)~\cite{depth_anything_v2} & $0.15 \pm 0.09$ \\
        RT-Mono-Depth~\cite{rtmonodepth} & $0.16 \pm 0.08$ \\
        RT-Mono-Depth (small)~\cite{rtmonodepth} & $0.11 \pm 0.09$ \\
        SGBM (filtered, GPU)~\cite{sgbm} & $1.73 \pm 0.26$ \\
        \hline
    \end{tabular}
    }
\end{table}

Table~\ref{tab:core_processing_latency} summarizes the average computational latency for each processing stage using the frequency-domain FFTW beamformer and SGBM for depth estimation. These values reflect the internal algorithmic cost of each module once a frame or audio chunk has entered the processing pipeline, and therefore exclude sensor-level buffering, GPU transfer delays, and operating-system scheduling jitter. Vision inference and depth estimation dominate the computational workload, while beamforming contributes less than 1~ms per audio chunk.

\begin{table}[ht]
    \centering
    \caption{Average processing computational latency per stage during full-system operation (mean~$\pm$~standard deviation for a 10 minute benchmarking run).}
    \label{tab:core_processing_latency}
    \resizebox{\columnwidth}{!}{%
    \begin{tabular}{lc}
        \hline
        \textbf{Processing Stage} & \textbf{Latency [ms]} \\
        \hline
        YOLO preprocessing & $1.25 \pm 0.12$ \\
        YOLO inference & $6.77 \pm 0.20$ \\
        YOLO postprocessing & $28.45 \pm 0.26$ \\
        Depth estimation (SGBM)* & $12.35 \pm 0.80$ \\
        Beamforming (FFTW) & $0.53 \pm 0.25$ \\
        Overlap-add & $0.0118 \pm 0.0169$ \\
        \hline
        Total & $49.36 \pm 0.91$ \\
        \hline
        \multicolumn{2}{l}{*\textit{With 1/2-size downsampled input image and CUDA acceleration.}}
    \end{tabular}
    }
\end{table}

To evaluate the responsiveness of the integrated system, we also measure the end-to-end latency using timestamps assigned at the moment a camera frame or audio chunk is captured. For each modality, the end-to-end latency is defined as
\begin{equation}
    t_{\text{e2e}} = t_{\text{proc,end}} - t_{\text{sensor,capture}},
\end{equation}
which includes camera and ALSA buffering, GPU scheduling delays, inter-thread queueing, and all computational stages reported in Table~\ref{tab:core_processing_latency}.

A 10-minute continuous benchmark yields a visual end-to-end latency of $64.1 \pm 10.6$~ms and an audio/beamforming latency of $35.6 \pm 4.2$~ms. The system latency is therefore dominated by the visual pipeline, resulting in a worst-path sensor-to-output delay of approximately 64~ms, comfortably below the 100~ms real-time target.

\begin{figure}[!htpb]
    \centering
    \includegraphics[width=0.8\linewidth]{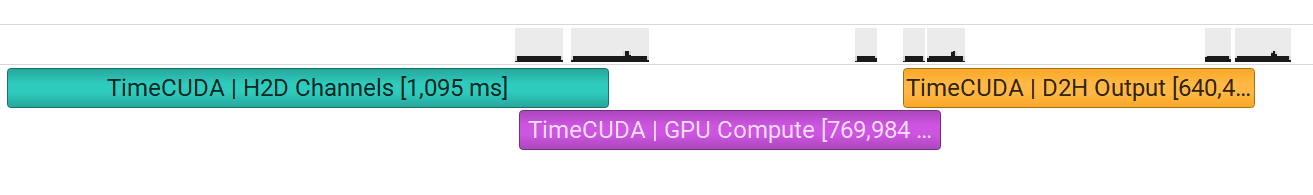}
    \caption{Arbitrary CUDA frequency-domain beamforming annotated range profiling with NVTX and Nsight Systems. A 10-second warm-up period was performed prior to measurement to ensure that initial memory and cuFFT allocations did not affect timing results.}
    \label{fig:profiling}
\end{figure}

Table~\ref{tab:beamformer_comparison} compares the average computation times for all beamforming variants. Time and frequency-domain implementations showed similar latency, with frequency-domain processing selected due to its improved scalability and flexibility for steering and filter design~\cite{Hamid2014}. The CUDA-based implementations did not provide practical performance gains. Profiling showed that data transfers between host and device memory consumed a substantial portion of the total execution time (see Fig.~\ref{fig:profiling}). Moreover, the execution times exhibited high variability across runs. Given the small chunk sizes required for real-time operation, performing beamforming entirely on the CPU proved more efficient.

\begin{table}[ht]
    \centering
    \caption{Comparison of the main processing blocks latency (mean~$\pm$~standard deviation).}
    \label{tab:beamformer_comparison}
    \begin{tabular}{lcc}
        \hline
        \textbf{Implementation} & \textbf{Domain} & \textbf{Latency [ms]} \\
        \hline
        CPU & Time & $0.43 \pm 0.25$ \\
        CPU (FFTW) & Frequency & $0.53 \pm 0.36$ \\
        CUDA & Time & $2.30 \pm 1.56$ \\
        CUDA (cuFFT) & Frequency & $2.05 \pm 1.18$ \\
        \hline
    \end{tabular}
\end{table}

Additional apparent delays stem from buffering and synchronization mechanisms within the multimodal pipeline rather than from computational processing. The effective beamforming delay of approximately 32~ms corresponds to the audio chunk duration (256 frames at 8~kHz), not to algorithmic latency. Similarly, the visual pipeline incurs a delay on the order of one frame interval due to capture buffering and inter-thread coordination, and this is already reflected in the end-to-end measurements above. These offsets are structural characteristics of the pipeline and can be reduced through tighter synchronization and shallower queues.

Overall, the profiling results indicate that the proposed implementation maintains stable real-time performance, with true sensor-to-output latency well below 100~ms. The integration of GPU-accelerated vision inference with optimized CPU-based frequency-domain beamforming offers a balanced compromise between accuracy, efficiency, and temporal consistency.

\subsection{Anechoic Environment}

\begin{figure*}[!htpb]
    \centering
    \includegraphics[width=\linewidth]{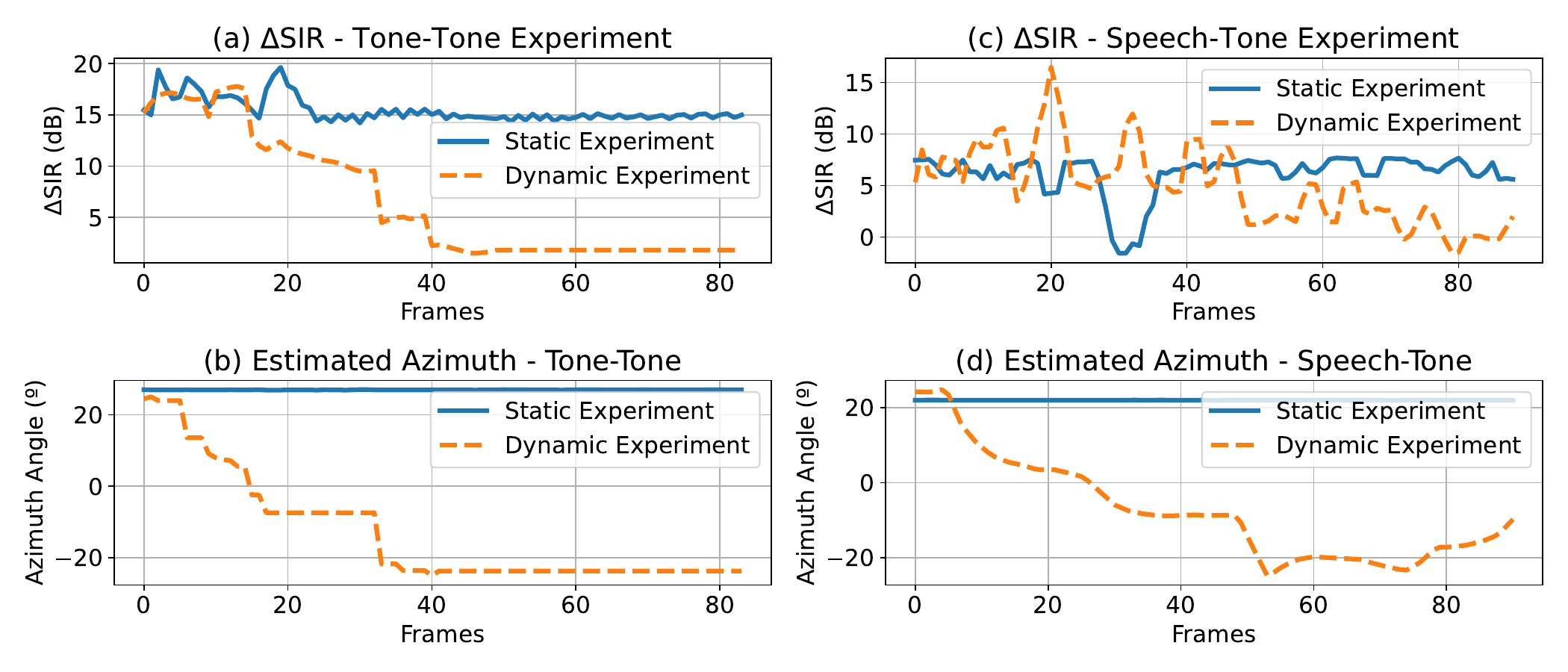}
    \caption{Results for the static and dynamic controlled environment experiments. Top row displays the SIR improvement for the (a) tone-tone experiment and for the (c) tone-speech experiment. The bottom row displays the estimated azimuth angle.}
    \label{fig:sir_anechoic}
\end{figure*}

\begin{figure}[ht]
    \centering
    \includegraphics[width=0.8\linewidth]{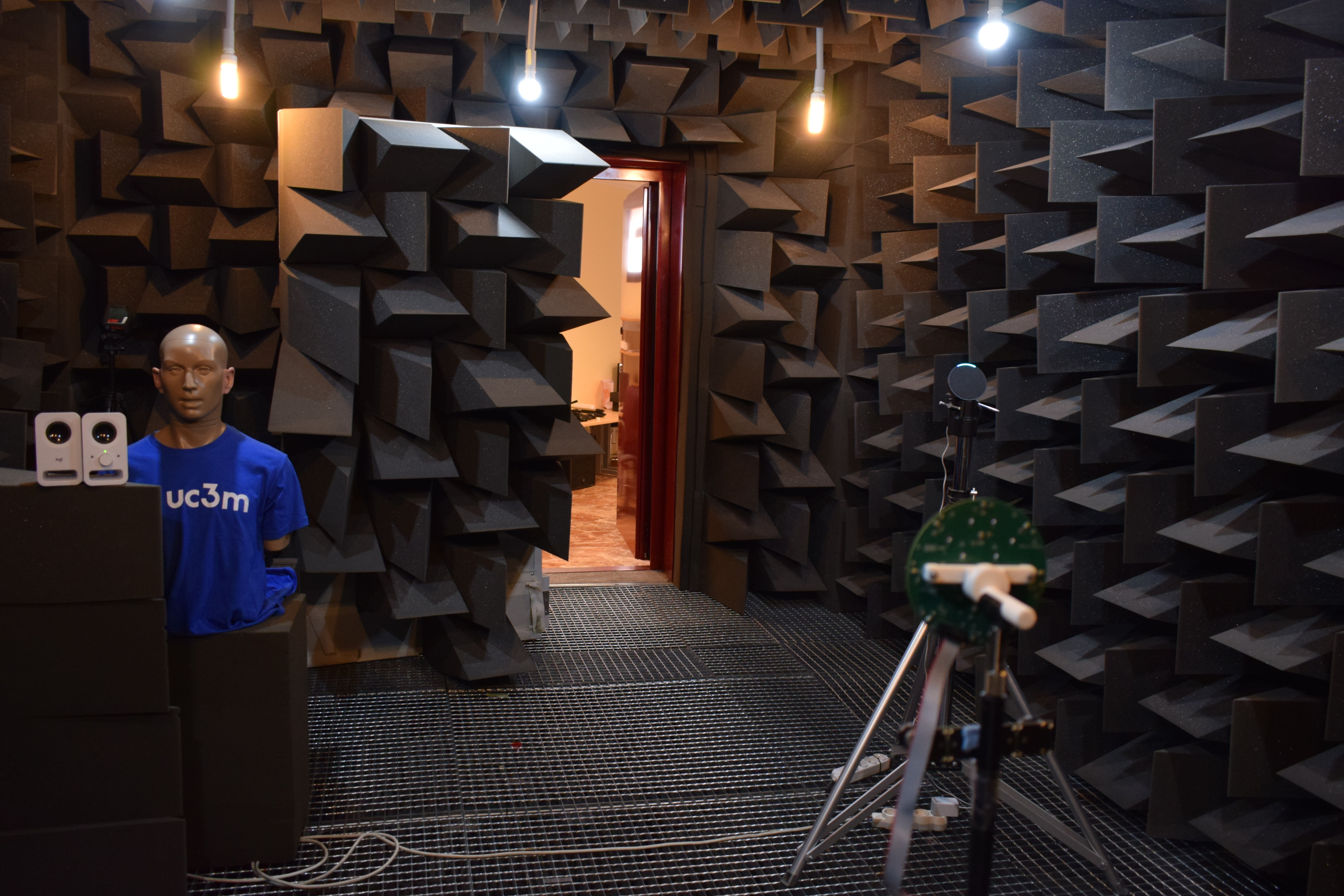}
    \caption{Experimental setup for the anechoic test.}
    \label{fig:static_anechoic_env}
\end{figure}

The first set of experiments was conducted inside an anechoic chamber to isolate the system from environmental noise and reflections. The setup included two household loudspeakers acting as sound sources and a dummy head representing the target (Fig.~\ref{fig:static_anechoic_env}). Both static and dynamic configurations were tested: in the dynamic case, one source was carried by a person walking at a normal pace toward the fixed source.

Two signal conditions were evaluated. In the first, the sources emitted pure tones at \SI{2}{\kilo\hertz} and \SI{3}{\kilo\hertz}. In the second, the \SI{3}{\kilo\hertz} tone was replaced by broadband speech noise derived from the TIMIT dataset~\cite{timit_dataset}. Playback levels for both loudspeakers were adjusted to comfortable, non-clipping values but were not SPL-calibrated, as our analysis focuses on the \emph{difference} between SIR measured on the beamformed and non-beamformed signals. This $\Delta\mathrm{SIR}$ metric is invariant to global volume scaling, provided that the signals remain above the noise floor and below saturation.

Separation performance was quantified using the signal-to-interference ratio (SIR). For the two-tone case, SIR was measured between the \SI{2}{\kilo\hertz} and \SI{3}{\kilo\hertz} components over time (Fig.~\ref{fig:sir_anechoic}a-b). Due to variability in source stability, the analysis focused on the SIR difference between beamformed and non-beamformed signals. Beamforming consistently improved SIR, confirming the system’s ability to enhance the target source. In dynamic conditions, SIR decreased as the moving source approached the fixed one, as both signals gradually fell within the same beam.

When the target emitted speech, SIR was computed according to Eq.~\ref{eq:SIR_speech}, where $\Delta \mathrm{SIR}$ represents beamforming gain and $\mathrm{PWR}_{\text{int}}$ is obtained by integrating the power spectral density over a \SI{100}{\hertz} band centered at the interference frequency. Results for static and dynamic speech conditions are shown in Fig.~\ref{fig:sir_anechoic}c-d.
\begin{align}
    \label{eq:SIR_speech}
    \Delta \mathrm{SIR} &= \mathrm{SIR}^{BF} - \mathrm{SIR}^{NBF} \ \left[\SI{}{\dB}\right], \\
    \mathrm{SIR} &= 10 \log_{10} \left( \frac{\mathrm{PWR}_{\text{speech}} - \mathrm{PWR}_{\text{int}}}{\mathrm{PWR}_{\text{int}}} \right) \left[\SI{}{\dB}\right] \nonumber
\end{align}

\subsection{Dynamic Environment}

\begin{figure}[ht]
    \centering
    \includegraphics[width=0.8\linewidth]{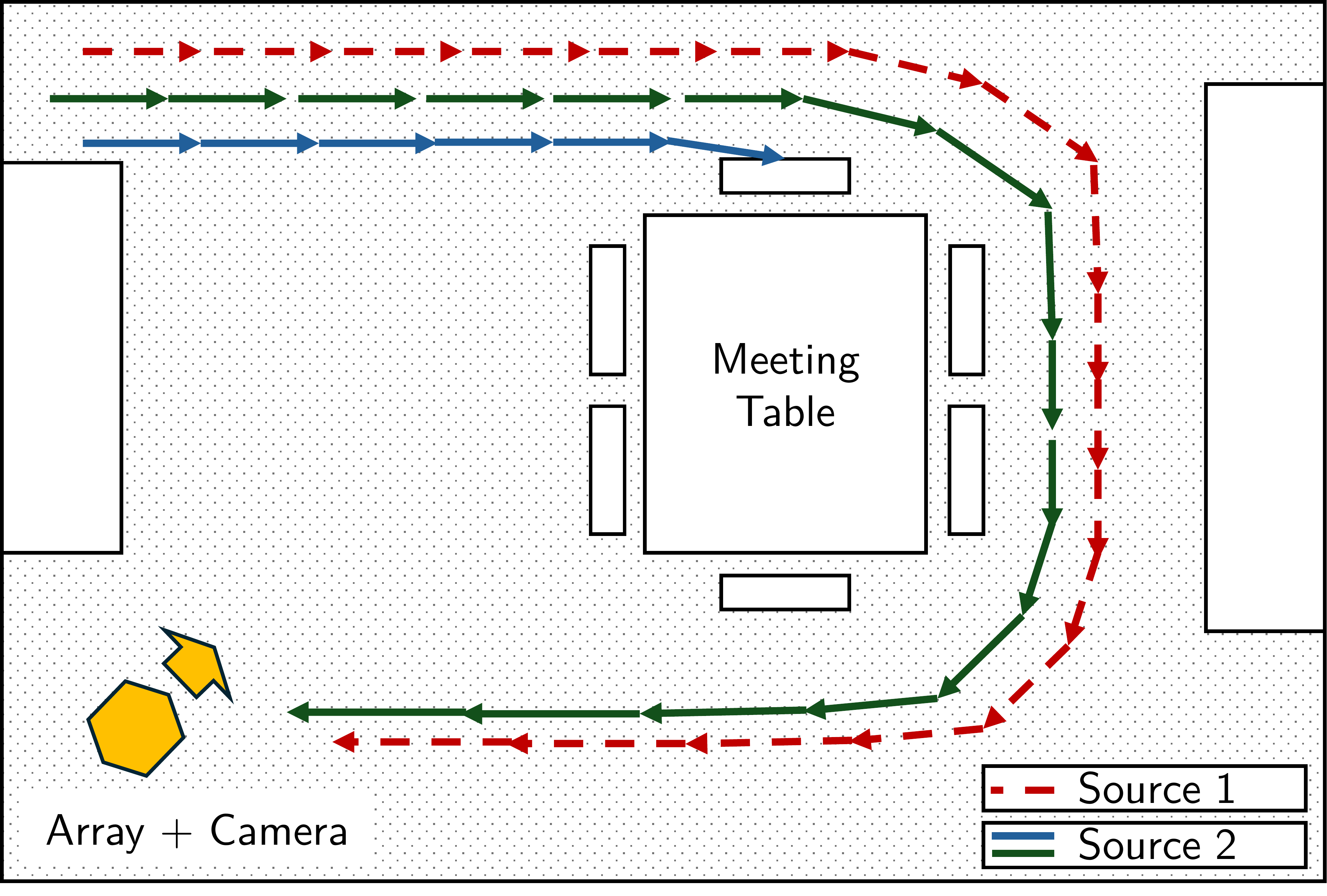}
    \caption{Configuration of the dynamic experiment. Top-down view of the recording area. Source 1 emitted a \SI{2}{\kHz} tone, while Source 2 alternated between a \SI{3}{\kHz} tone and white noise (in separate runs).}
    \label{fig:exp2setup}
\end{figure}

To extend the evaluation beyond the controlled chamber, experiments were also conducted in a typical room environment resembling a meeting scenario, with a central table and reflective surfaces (Fig.~\ref{fig:exp2setup}). Two participants each carried a mobile phone acting as a moving sound source, while the recording system remained fixed. Two participants each carried a mobile phone acting as a moving sound source, while the recording system remained fixed. The output levels of both phones were set to comfortable, non-clipping values without SPL calibration. As in the anechoic experiments, we report relative SIR improvements ($\Delta\mathrm{SIR} = \mathrm{SIR}^{BF} - \mathrm{SIR}^{NBF}$), so the results are robust to global changes in playback volume under the same recording conditions.

\begin{figure}[ht]
    \centering
    \includegraphics[width=0.8\linewidth]{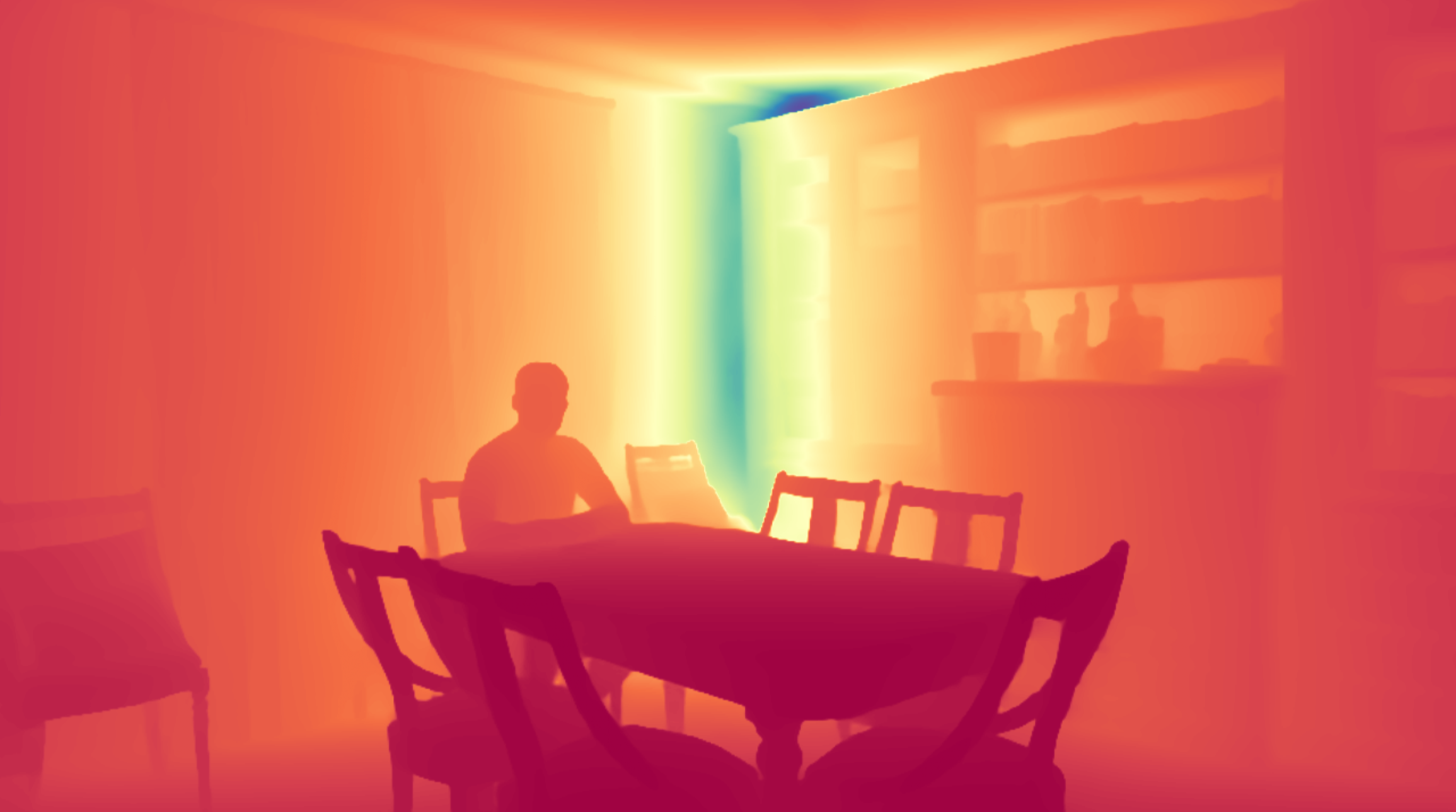}
    \caption{Depth map of the experiment scene (obtained with \cite{depth_anything_v2}).}
    \label{fig:exp2depth}
\end{figure}

The participants walked sequentially rather than in parallel: Source~2 began moving first, and Source~1 started walking once Source~2 had advanced roughly \SI{1}{\meter}. Both then walked at a similar natural pace, resulting in a non-fixed, organically varying separation. Figure~\ref{fig:exp2depth} illustrates the setup and corresponding depth map of the scene.

Two conditions were tested: one with pure tones at \SI{2}{\kilo\hertz} and \SI{3}{\kilo\hertz}, and another in which the \SI{3}{\kilo\hertz} tone was replaced by broadband white noise. Although mobile phone loudspeakers exhibit limited directivity and stability~\cite{phonespeaker}, this had minimal impact, as analysis relied on relative SIR improvements. Moreover, using phones as sources reflects realistic interference scenarios common in daily environments.

\begin{figure*}
    \centering
    \includegraphics[width=0.41\linewidth]{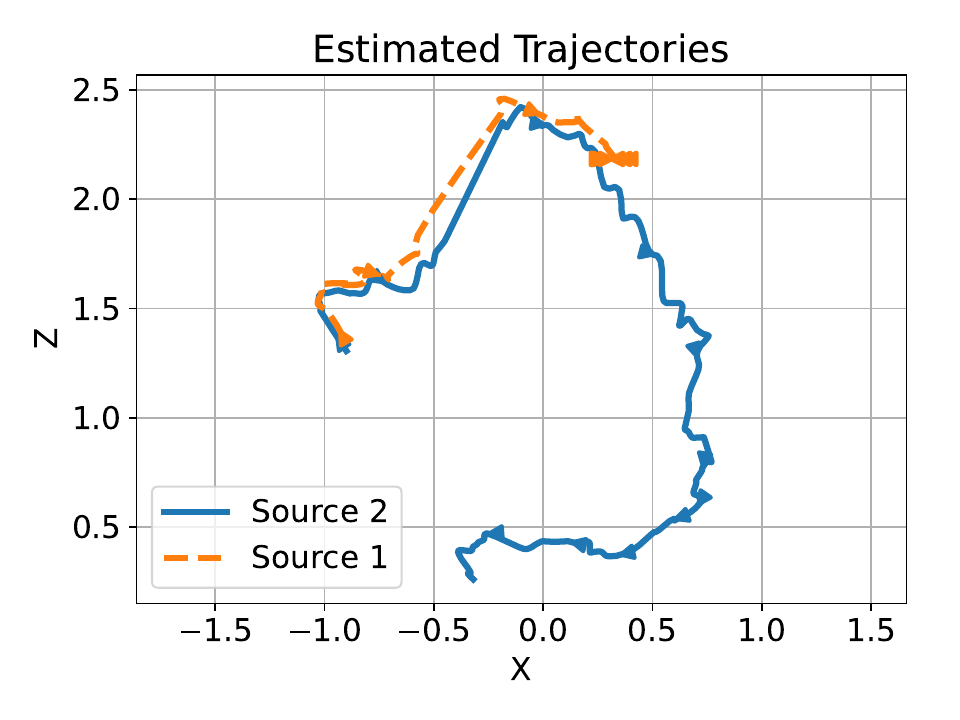}
    \hspace{2em}
    \includegraphics[width=0.48\linewidth]{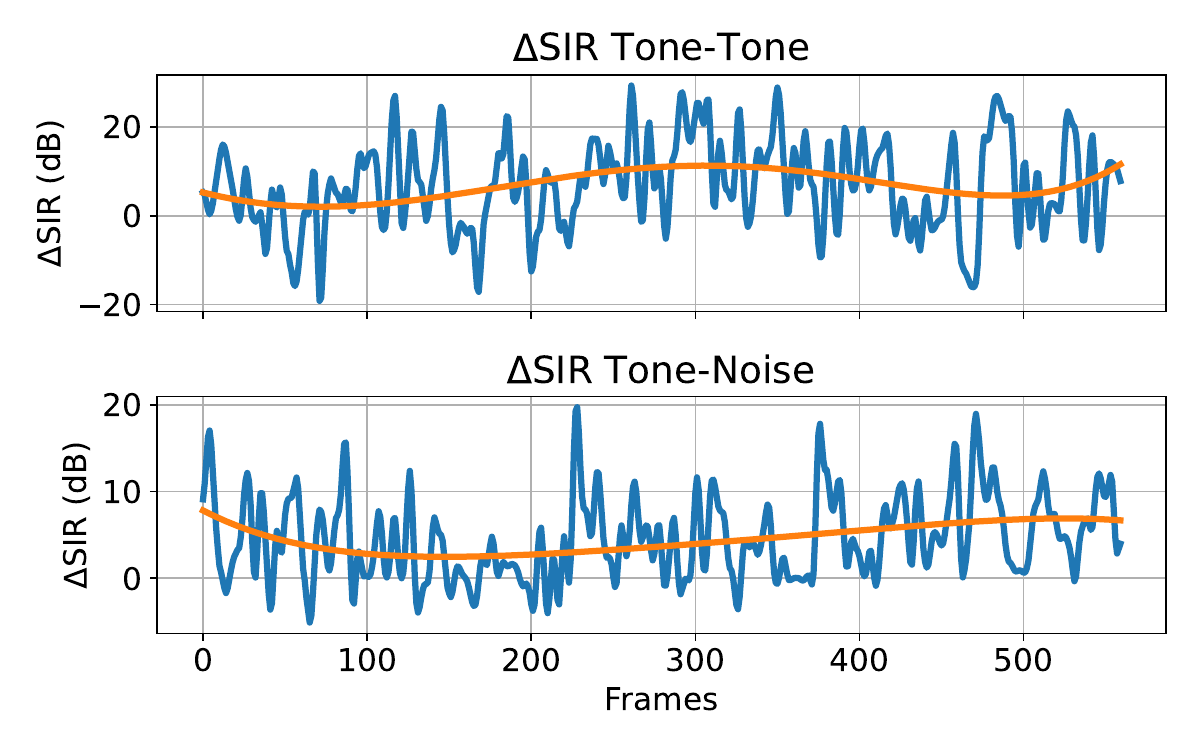}
    \caption{Left shows the estimated trajectories of the sources in the dynamic experiment. Source 1 remains seated as shown in Fig.~\ref{fig:exp2depth}, while Source 2 completes its trajectory. Right shows the SIR improvement obtained in the dynamic environment experiments. A fitted curve of order five is included for easier interpretation.}
    \label{fig:room_experiment}
\end{figure*}

Compared to the anechoic setup, the reflective room and variable phone acoustics introduced greater measurement variability. To mitigate this effect, speech was replaced with white noise in the broadband condition. Figure~\ref{fig:room_experiment} (right) shows the SIR improvement between beamformed and non-beamformed signals for both the two-tone and tone-plus-noise cases. As shown, both cases benefited from beamforming, with generally positive $\Delta$SIR values indicating effective interference suppression despite environmental variability.

Finally, Fig.~\ref{fig:room_experiment} (left) presents the trajectories estimated by the vision module. Despite partial occlusions, the deep learning-based detection successfully tracked both sources independently, confirming the system’s robustness under realistic dynamic conditions. 

\section{Conclusion}

This work presented an embedded real-time system that integrates visual depth estimation with acoustic beamforming for spatially selective audio capture. By combining camera-based localization with frequency-domain delay-and-sum processing, the system achieves effective sound source separation and adaptive directional filtering in both static and dynamic environments. Experimental validation demonstrated consistent improvements in signal-to-interference ratio, confirming that vision-guided beam steering significantly enhances isolation of desired sources, particularly when spatial separation is maintained.

Filtered SGBM was selected for our experiments as it provided the most stable and accurate depth estimation across varying lighting and texture conditions, ensuring reliable evaluation even in less common environments such as anechoic chambers. While other, more scene-dependent methods demonstrated higher processing speed and lower latency, their performance tended to vary in controlled or acoustically unique spaces. These findings highlight the balance between robustness and efficiency required in embedded multimodal systems. A monocular approach, more suitable for typical indoor scenarios, is identified as a promising direction for future work. Furthermore, the modular software architecture and CPU–GPU co-processing strategy enabled sustained real-time performance with sufficiently low latency, confirming the suitability of the proposed design for resource-constrained platforms.

Overall, this study demonstrates that integrating visual spatial awareness with frequency-domain beamforming offers a practical and scalable solution for intelligent audio capture in real-world scenarios such as teleconferencing, assistive devices, and human–robot interaction. Future work will explore adaptive beamforming strategies and more advanced depth fusion techniques to further improve performance in complex, multi-source acoustic environments.

\section*{Declarations}
\label{sec:dec}

\subsection*{\textbf{Ethical Approval}}

Not applicable.

\subsection*{\textbf{Competing interests}}

The authors declare that they have no known competing financial interests or personal relationships that could have appeared to influence the work reported in this paper.

\subsection*{\textbf{Authors' contributions}}

All authors contributed equally to this work.

\subsection*{\textbf{Funding}}

This work has been supported by the Spanish Ministry of Science and Innovation under projects PID2023-148671OB-I00, PID2022-137048OB-C41 and PID2022-137048OA-C43 funded by MCIN/AEI/10.13039/501100011033 and by ``ERDF A way of making Europe" as well as TED2021-131401B-C21 and TED2021-131401A-C22. This action has been also funded through the R\&D activities program with reference TEC-2024/COM-360 and acronym DISCO6G-CM granted by the Community of Madrid through the General Directorate of Research and Technological Innovation through Order 2402/2024".

\subsection*{\textbf{Availability of data and materials}}

No additional data or materials available.

\bibliography{ref}
\bibliographystyle{icml2025}


\end{document}